# Amino acid preference against beta sheet through allowing backbone hydration enabled by the presence of cation

John N. Sharley, University of Adelaide.

john.sharley@pobox.com

## Table of Contents

## 1   Abstract

It is known that steric blocking by peptide sidechains of hydrogen bonding, HB, between water and peptide groups, PGs, in beta sheets accords with an amino acid intrinsic beta sheet preference [1]. The present observations with Quantum Molecular Dynamics, QMD, simulation with Quantum Mechanical, QM, treatment of every water molecule solvating a beta sheet that would be transient in nature suggest that this steric blocking is not applicable in a hydrophobic region unless a cation is present, so that the amino acid beta sheet preference due to this steric blocking is only effective in the presence of a cation. We observed backbone hydration in a polyalanine and to a lesser extent polyvaline alpha helix without a cation being present, but a cation could increase the strength of these HBs.

Parallel beta sheets have a greater tendency than antiparallel beta sheets of equivalent small size to retain regular structure in solvated QMD, and a 4 strand 4 inter-PG HB chain parallel beta sheet was used. Stability was reinforced by one surface being polyvaline, which buttressed the opposite surface which was used for experimentation. A single $Ca^{2+}$ ion was used for investigation of individual binding events rather than bulk properties. No direct binding between $Ca^{2+}$ and the PG oxygen was observed in these simulations, but perhaps it occurs at longer time scales as the transient beta sheet unfolds.

When linear scaling QMD methods that are accurate [2, 3] for peptide resonance, Resonance-Assisted Hydrogen Bonding [4, 5], RAHB, and the properties of water become available, more extensive experiments having multiple ions of multiple types could be performed at acceptable computational cost. It important that such investigations be performed on protein secondary structures rather than model amides so that sidechain limitation of backbone hydration and hence intrinsic amino acid propensity is captured.

## 2 Introduction

### 2.1 Alpha helix preferring amino acid residues in a beta sheet

Sidechain steric blocking of backbone hydration is known to correlate with amino-acid strand preference [1]. It is also known that strand preferring amino acids have minimum steric conflict with the backbone in that preferred conformation [6], but this does not provide a mechanism for disfavour of alpha helix preferring residues in beta sheets whereas steric blocking of backbone hydration does. The greatest preference for alpha helices is that of alanine [7], though the backbones of polyalanine alpha helices are hydrated, with alpha helical preference not correlating with sidechain blocking of backbone hydration in alpha helices [1]. We used a beta sheet with a polyalanine surface, with the other surface being polyvaline for stabilization of the sheet. Parallel beta sheets are more stable than antiparallel beta sheets when the sheets are small, and a 4 chain 4 inter-PG HB parallel beta sheet was used here. Valine has the largest preference for beta sheets [7], and in parallel beta sheets, valine sidechains of a polyvaline surface can interlock in a regular manner. The strands were acetyl and N-methyl capped rather than left charged, to more closely resemble unbroken chains (Figure 1).

Hydration and destabilization of a beta sheet backbone was studied at the level of individual HB events rather than at the level of bulk behaviour which does not necessarily illuminate basic molecular mechanism. The beta sheet studied would be transient if it occurred in nature at all, with its occurrence constituting a transient misfolding of the involved residues. Illustration of the patterns of HB contributing to the correction of this transient misfolding was sought in the following experiments.

## 2.2 Cation interactions with protein backbone oxygen

Algaer and van der Vegt [8] note that the chemical environment of the backbone amide should be considered in studying Hofmeister effects [9]. To this end, we investigated water mediated cation interactions with backbone amides in a beta sheet that would be transient in nature.

Okur *et al*. [10] found that the direct contact and solvent-separated binding for cations and amide oxygen in d-butyramide was extremely weak at biological ion concentrations, and the ordering of affinity is $Ca^{2+}$ > $Mg^{2+}$ > $Li^+$. Therefore, we used $Ca^{2+}$ for this study to best observe cation interactions with the backbone.

## 2.3 Quantum molecular dynamics with quantum mechanical treatment of every water molecule

Okur *et al*. [10] state that "simulations that find tight associations between metal cations and the carbonyl oxygen of amides are not consistent with the spectroscopic data". This motivates the use of fully QM rather than classical simulations.

Our QM investigations of variation in amide resonance due to electrostatic field with component parallel or antiparallel to the amide C-N bond and due to RAHB demonstrate that the variation in amide resonance and hence charge distribution in the amide group is considerable. Classical calculation has no account of variation of amide resonance. Since the backbone amides studied were already participating in secondary structure RAHB chains and further we are introducing ions and hence electrostatic field into the vicinity of the backbone group, classical methods are unsuitable. Against this indispensable advantage of QM methods, there is our own finding that established Density Functional Theory, DFT, methods are undesirable for calculation of amide resonance when the amide carbonyl is engaged in torsional hyperconjugative interactions that in particular occurs in parallel and antiparallel beta sheets [11]. However, pending the availability of methods that accurately model amide resonance and scale to QM-handled explicitly solvated beta sheets, we are obliged to use these DFT methods.

The computational expense of these simulations with available methods prohibits a comprehensive survey. Such a survey can be undertaken upon the availability of QMD methods that scale linearly in runtime and memory use with both protein atom count and solvent atom count and provide a good account of amide resonance, RAHB and the properties of water. Fragment methods [12] need to be qualified as accurately capturing RAHB in secondary structure backbone amide chains. If fragment boundaries bisect peptide bonds, particular care needs to be taken that amide resonance varies accurately with the electrostatic field component that is parallel or antiparallel to the amide C-N bond [2] as well as with RAHB-induced variations. Surveys involving greater depth of solvent and longer simulated duration that could be undertaken with more accurate and better scaling methods are described in Section 7 (Future work).

## 3  Methods

The quantum chemistry package TeraChem 1.5K [13-16] was used in these large atom count QMD experiments. This version of TeraChem is cubic scaling at the atom count used, proceeding further up the cubic curve than other programs by efficient utilization of Graphical Processor Units [17]. Explicit solvent is used, with quantum mechanical treatment of all solvent molecules.

While the protein is fully solvated, the solvation shell is thin in classical simulation terms, being 7 angstroms, but is similar to the 5 angstrom minimum solvation shell recommended for best maintenance of the HOMO-LUMO gap in DFT calculations of protein electronic structure [18]. In the molecularly crowded intra-cellular environment, it will often happen that solvent between proteins is sparse, though there is a water/vacuum interface in the present simulations.

Full Hartree-Fock exchange [19] at long range is recommended to minimize a charge transfer inaccuracy arising from decreasing HOMO-LUMO gap of edge waters with oxygen not participating in HB [20]. We used LC-wPBE [21], denoted wpbe in TeraChem, for its tractability of convergence. Isborn *et al*. [20] give w=0.26 as Koopman-optimal [22] for this method, but they used 0.2 and we used 0.4. D3 empirical correction [16] is not available for this method.

No geometry constraints were applied other than protein atoms fixed for early steps in some experiments while the TIP3P solvent shell added by the Chimera [23] Solvate function took on angles, bond lengths and density suitable for QMD. The course of the QMD simulation was unguided by Ab Initio Steered Molecular Dynamics [24] or other means. Spherical boundary conditions 'mdbc spherical', constant density 'md_density 1.0' and Langevin dynamics 'thermostat langevin' were set. The Langevin temperature damping time 'lnvtime' [25] was varied during equilibration as given in the notes for each experiment. The indicator of equilibration used is approximate convergence to the target temperature, which was set at 310.15 Kelvin.

Focussing on cations has a computational benefit in that there is less need for diffuse functions [26], in contrast to anions which have more diffuse electron distribution. Diffuse functions are more challenging for convergence of quantum chemistry methods, and computational speed is diminished by their use. Also, in a QM-handled explicitly solvated simulation, electrons tend to be less diffuse than in the gas phase, reducing the need for diffuse functions.

We are presently obliged to use basis sets of a size and type we demonstrated to contribute to inaccuracy in modelling amide resonance and hence secondary structure RAHB [11].

# 4 Results

## 4.1 Preparation

Chimera's Solvate function was used to add a 7.0 angstrom shell of TIP3BOX waters to a 4 strand 4 inter-PG HB chain parallel beta sheet with valines on one side to stabilize the sheet and alanines on the other side for the experimental surface (Figure 1) which had been geometry optimized at wpbe(rc_w=0.4)/6-31g** before solvation. 300 steps (frames) of QMD was applied with the coordinates of heavy atoms fixed then 200 steps with all atoms free, with all steps at a Langevin thermostat damping time parameter value lnvtime=10 fs.

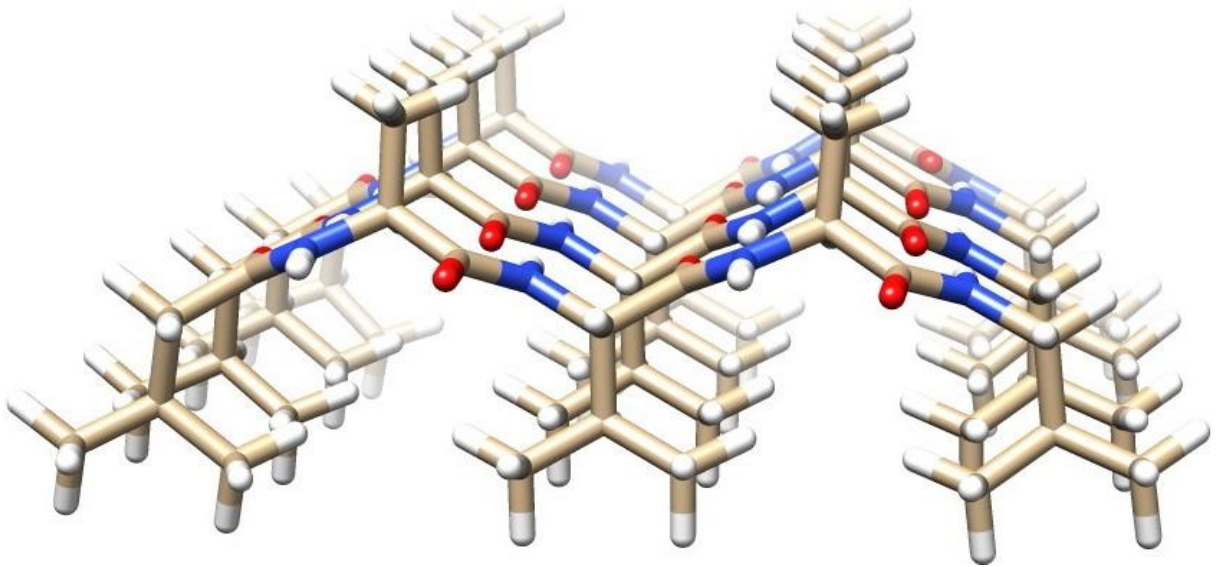

Figure 1. 4-Beta Strand 4-HB Chain Parallel Beta Sheet with Ala on Experimental Surface and Val on Other Surface after Unconstrained Optimization at wpbe(rc_w=0.4)/6-31g** Prior to Solvation

## 4.2 Experiment 1302

A water (oxygen serial number 436) with coordinates 3.359 angstroms from an Ala CB on one cross-strand row and 5.271 angstroms from the Ala CB on the other such row on the same strand on the same surface of the beta sheet was replaced with a $Ca^{2+}$ atom at the oxygen coordinates and a new unconstrained MD run (1302) was started from that frame. 150 QMD steps with lnvtime=10 fs then 650 steps with lnvtime=100 fs were applied.

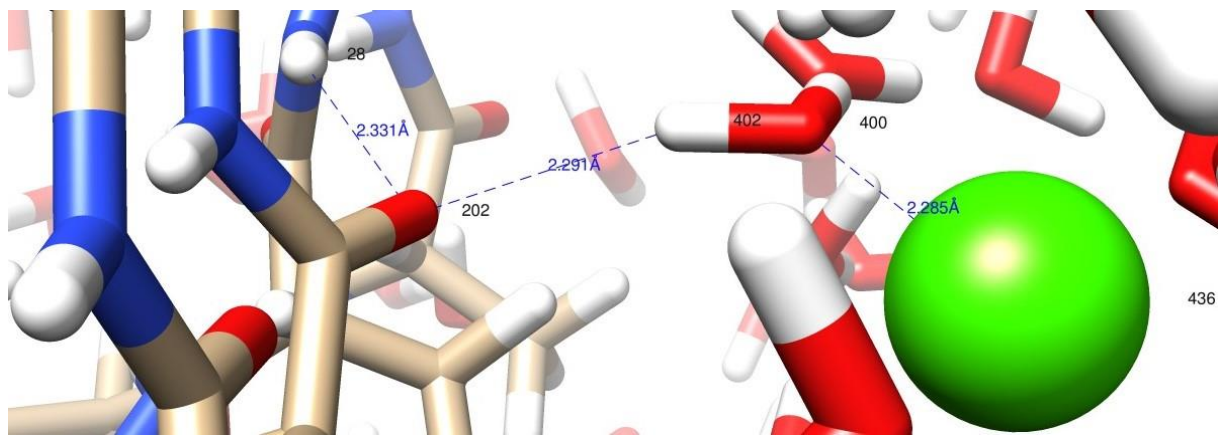

Figure 2. Backbone Hydration in Final Frame of Experiment 1302

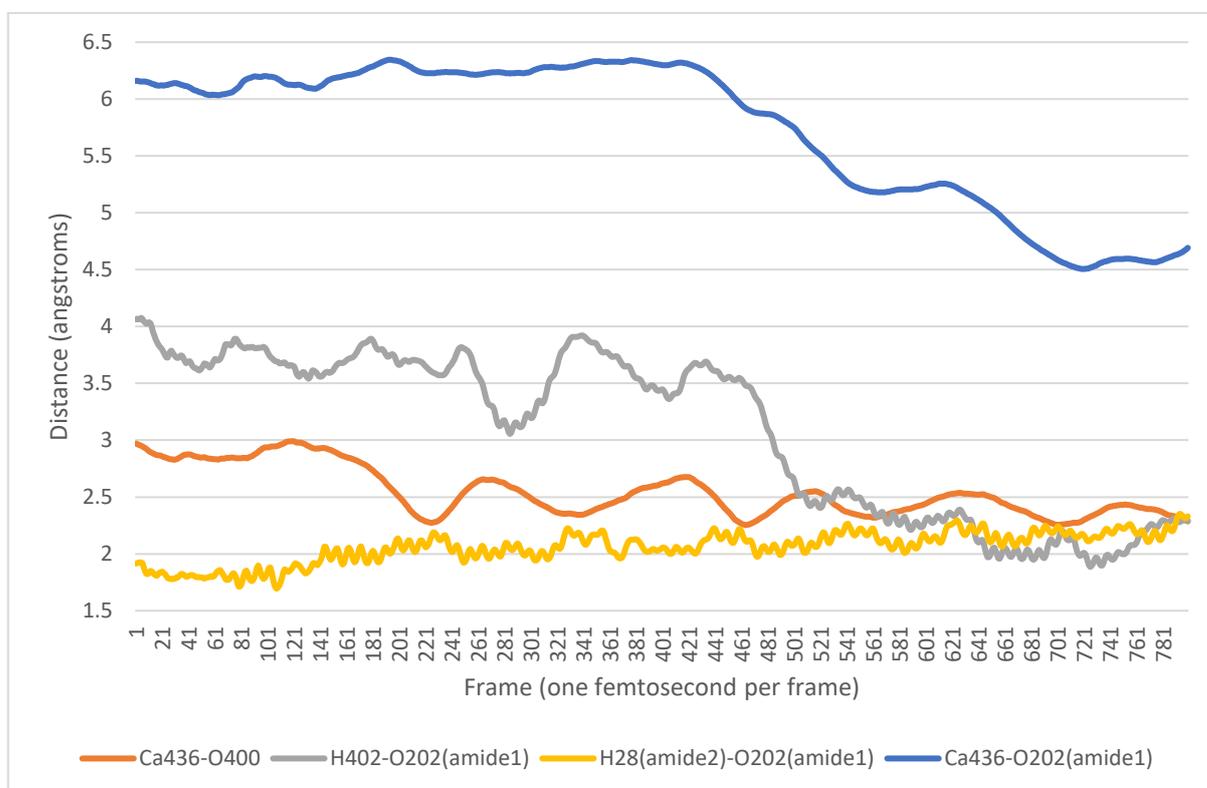

Figure 3. Selected Bond Lengths in Experiment 1302. Atom Ids Shown in Figure 2

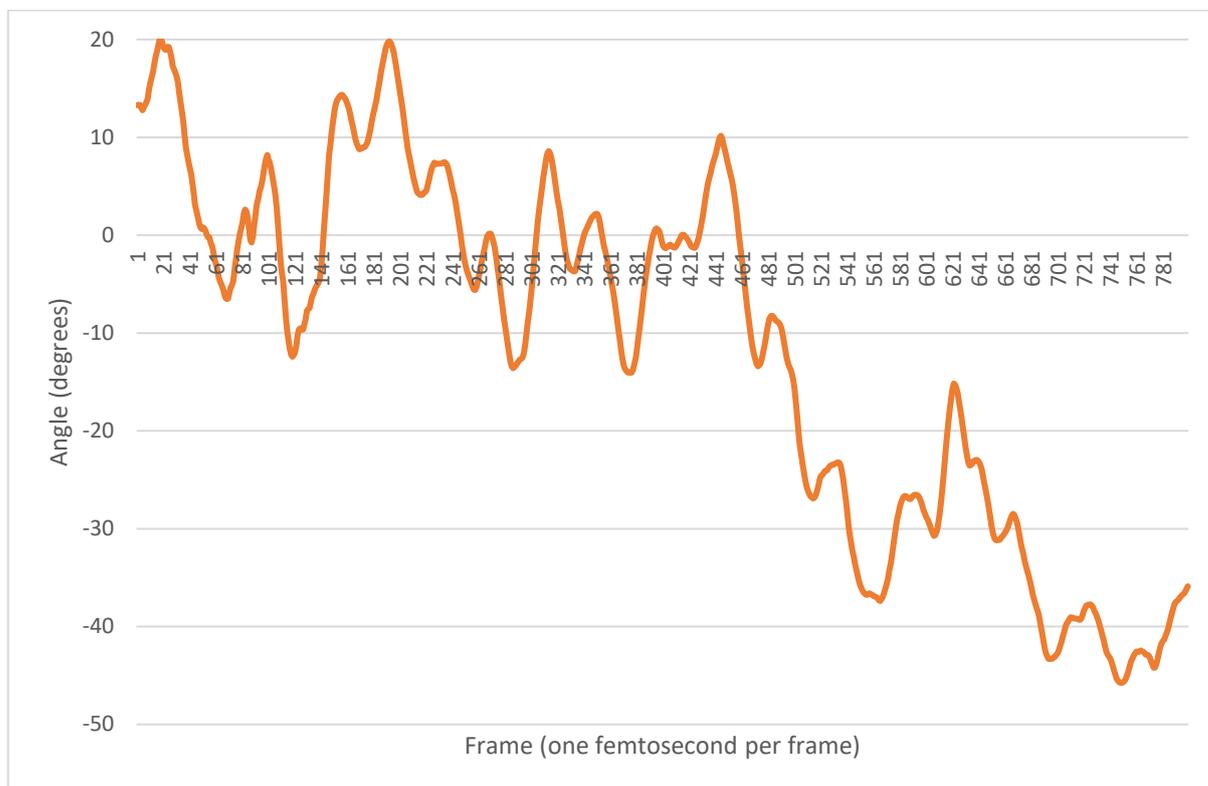

Figure 4. Angle of O202-H28 from O202, C200, N201 Normal Minus 90.0 in Experiment 1302

### 4.3 Experiment 1303

Starting with the final frame resulting from the common preparation (1301) and removing a water (oxygen serial number 391) with coordinates 4.915 angstroms from an Ala CB and 5.561 angstroms from another Ala CB on the same cross-strand row and placing a $Ca^{2+}$ atom at the oxygen coordinates, a new unconstrained QMD run (1303) was started. 150 QMD steps with lnvtime=10 fs then 650 steps with lnvtime=100 fs were then applied.

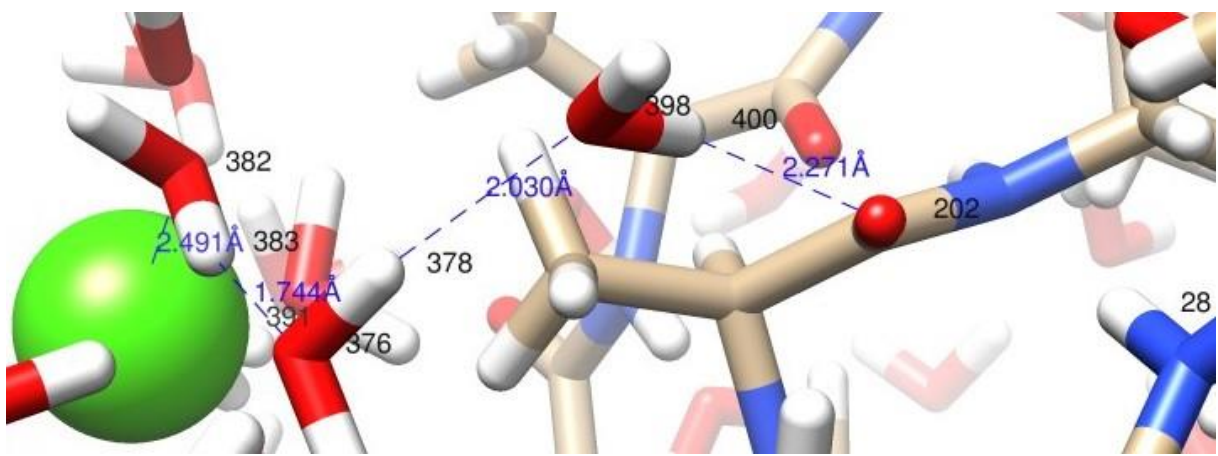

Figure 5. Backbone Hydration in Final Frame of Experiment 1303

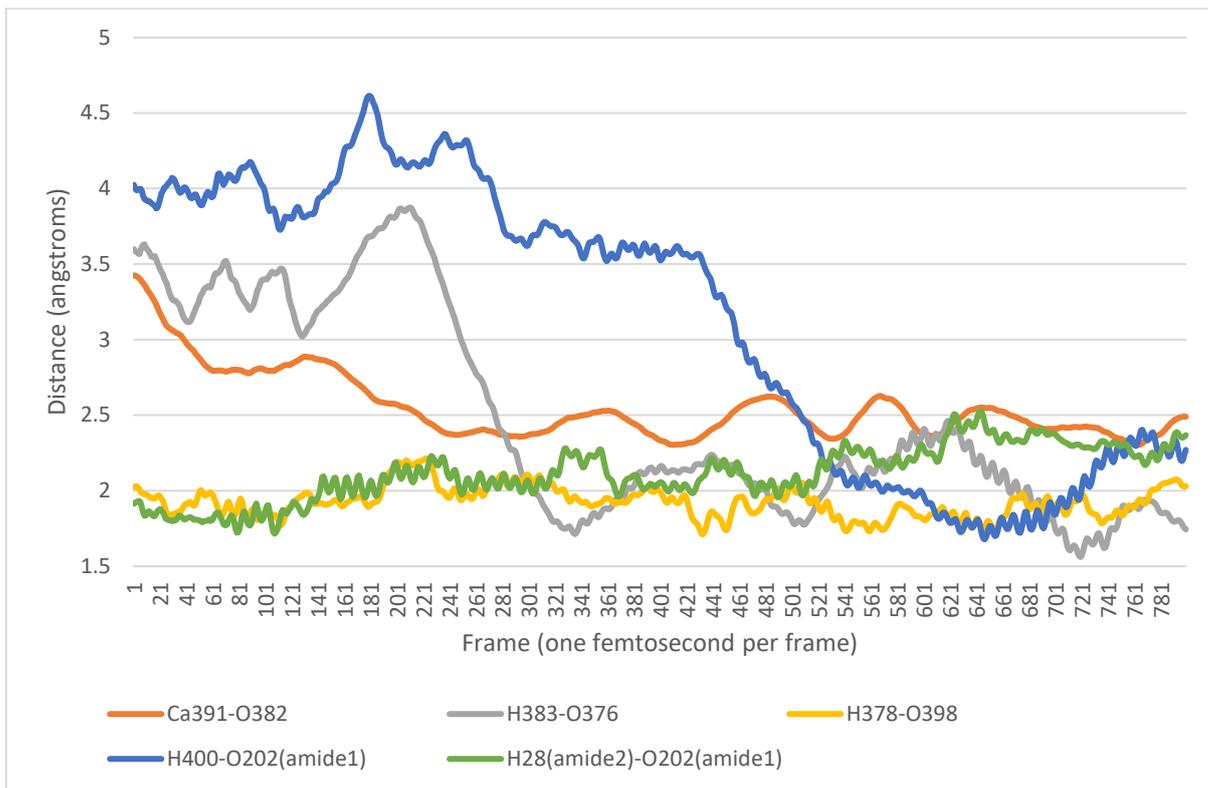

Figure 6. Selected Bond Lengths in Experiment 1303. Atom Ids Shown in Figure 5

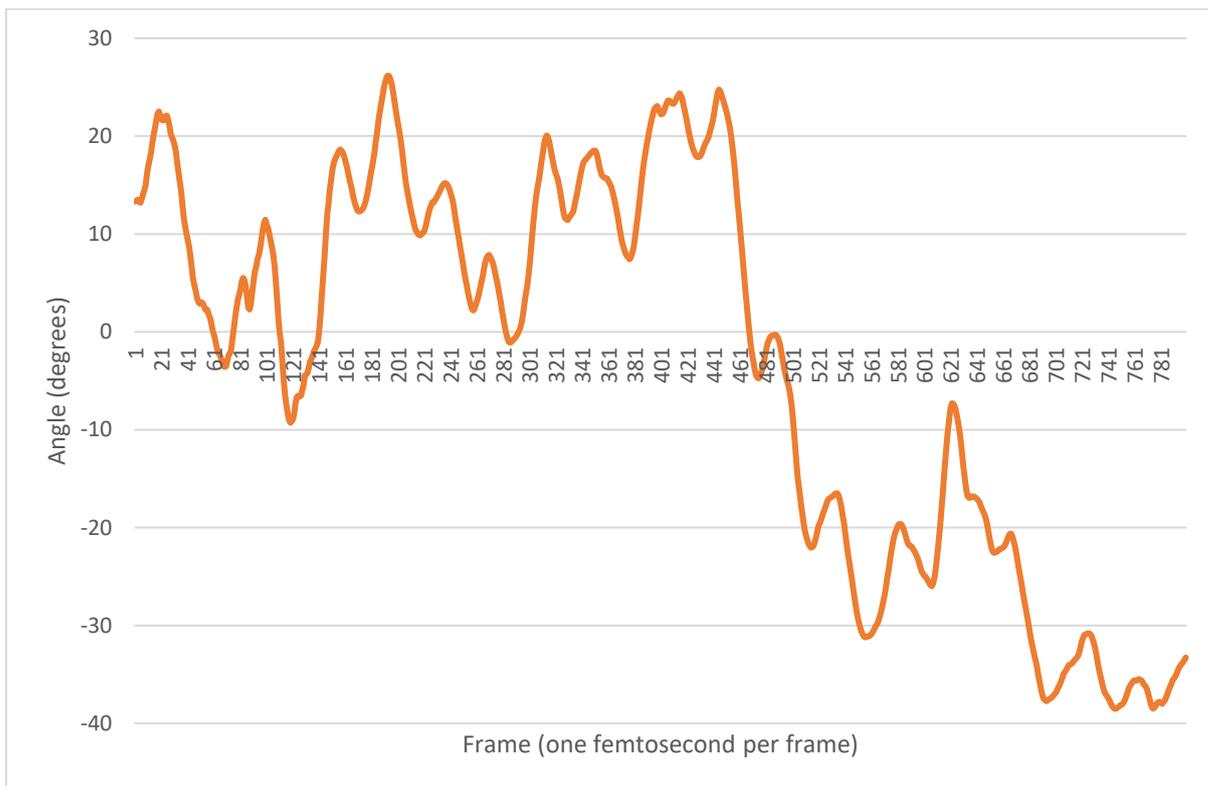

Figure 7. Angle of O202-H28 from O202, C200, N201 Normal Minus 90.0 in Experiment 1303

## 5 Discussion

### 5.1 HB networks of water

In both experiment 1302 and 1303, temperature is still equilibrating up to frame 440.

The experimental surface was between the two rows of alanines, since the backbone exposure outside of the alanine rows but still on that surface of the beta sheet was artificially large due to the strands not continuing beyond the final valines on the opposite surface. The two examples above, Figure 2 and Figure 5, show hydration between the rows of alanines, that is, on the experimental surface. Two examples of hydration outside the rows of alanines are shown in Ap1:Figure 8 and Ap2:Figure 11.

Figure 2 shows a single water connecting $Ca^{2+}$ and an interior PG. Figure 5 shows a chain of three water molecules connecting $Ca^{2+}$ and an interior PG. Ap1:Figure 8 shows a 3-deep network of waters connecting $Ca^{2+}$ and an interior PG outside the experimental surface. Ap2:Figure 11 shows a 1-deep network of waters connecting $Ca^{2+}$ and an interior PG outside the experimental surface. Perhaps connection of $Ca^{2+}$ to interior PGs between the rows of alanines requires a chain rather than a network of waters due to the steric constraints given by the alanine sidechains.

The distances between atoms as they become part of a HB network connecting $Ca^{2+}$ and PG oxygen are shown in Figure 3, Figure 6, Ap1:Figure 9 and Ap2:Figure 12.

Figure 4, Figure 7, Ap1:Figure 10 and Ap2:Figure 13 show the extent of variation of the angle between the PG normal and the inter-PG HB of its oxygen. At these very large angles from PG co-planarity, the RAHB of the inter-PG HB chain will be reduced as will the stability of the beta sheet. Whether the sheet unfolds or not will depend on the extent to which the geometry of all of the inter-PG HB chains in the sheet is ultimately de-optimized by this backbone hydration event or other such events.

In the presence of a single $Ca^{2+}$ ion, the backbone of the beta sheet is hydrated via the polyalanine surface. This supports a view that a $Ca^{2+}$ can rupture a hydrophobic hydration shell to enable intrinsic amino acid beta sheet preference. It cannot be concluded that water without an ion will never hydrate the backbone of the experimental surface, for the simulated time is short (0.8 picosecond, including temperature equilibration during which these HB networks do not form), but it is apparent that over the timecourse of the simulation, the backbone is more strongly hydrated in the presence of $Ca^{2+}$.

### 5.2 Subsequent to rupture of a transient beta sheet

If a transient beta sheet has been ruptured by cation-mediated backbone hydration, what happens next? Model amide studies are not revealing of the manner of backbone hydration of a ruptured beta sheet. Will not the HB chains and networks between the cation and the PG oxygen become stronger due to better access by water to the PG oxygen which is no longer in an inter-PG HB chain? This better access may include improved access to the PG oxygen p-type lone pair, increasing HB strength [3]. It is likely that quite a number of PG oxygens will be solvent-exposed after rupture of a beta sheet. The presence of cation makes peptide-water HB more favourable than inter-peptide HB. By what process are the cations eliminated so that the protein can fold? Perhaps in time the cations will associate with anions and drift away. Perhaps ongoing conformational change due to hydrophobia eventually dislodges these cations from the often hydrophobic residues of a beta sheet. Perhaps a number of solvent exposed PG oxygens in an otherwise hydrophobic surface is a target for molecular chaperone [27] action, whether the PG oxygens are no longer associated with cation, are indirectly or directly bound to cation.

Cations associating with PG oxygen that were not in an inter-PG HB chain would be limiting of folding progress. The prospects for a timely fold would be improved by protecting surface PG oxygen that would be buried in the correctly folded protein. Molecular chaperones might either protect such PG oxygens, or actively restructure the protein.

Perhaps a function of high $Ca^{2+}$ concentration such as in the endoplasmic reticulum, ER, is to test the folding of proteins in that compartment. Misfolded secondary structures of proteins of that compartment become unstable in the presence of $Ca^{2+}$ by the means described and folding chaperones then identify folding work to be done by the state of the protein after failing the test of folding.

### 5.3 Hofmeister effects

Perhaps how well other residues complement a given residue in obstructing the formation of HB networks between ions and PGs is dependent on the concentration of ions and temperature. If so, Hofmeister effects [28-30] must be studied in the presence of this sidechain complementation rather than in model amides. A structural account of this complementation such as can be given by QMD is desirable. Perhaps in aspects of Hofmeister effects, the mechanism that makes misfolded beta sheets more energetically unfavourable is dysregulated by non-physiological conditions. This might extend to sequence-dependent dehydrated sections of alpha helical backbone.

Hydrogen bonded networks connecting cations with PG oxygens are not necessarily diminished in polarization or charge transfer by the presence of anions. A water molecule might be a node of a HB network which includes both cation(s) and anion(s) such that its polarization or charge transfer is increased by interactions of its oxygen indirectly due to cation and also by interaction of one its O-H antibonding orbitals indirectly due to anion. Its free O-H antibonding orbital could then be indirectly associated with increased polarization or charge transfer from PG oxygen lone pairs.

Shi and Wang [31] suggest that cation and anion are in contact at high concentrations rather than being separately solvated. There are two quite opposing views arising from this that might be considered. The first is that this is inhibitory of HB networks connecting cations with PG oxygen. The second is that a water molecule with oxygen within charge transfer distance of the cation and a hydrogen within charge transfer distance of the anion may be yet more polarized than in the presence of cation alone, and the other hydrogen of the water molecule could start a more polarized HB chain to the backbone oxygen. Since there are likely to be multiple such waters at charge transfer distance with both cation and anion, they may collectively start a HB network to backbone oxygen which is more polarized than those from cation without anion. More polarized HB networks could be expected to rupture a hydrophobic hydration shell and access backbone oxygen where this would not occur at lower concentrations.

Another possibility, which assumes separability of ions, for the emergence of Hofmeister effects is that the electrostatic field with component parallel or antiparallel to the PG C-N bonds disturbs PG resonance and secondary structure RAHB chains [2] at high ionic concentrations, with the electrostatic field fluctuating with the particular arrangement of ions. A momentary clustering of anions near one surface of a protein and of cations on an opposed surface may be destabilizing of the electrostatic field in the protein between these surfaces and of the PG resonances and hence secondary structure RAHB. Brief electrostatic destabilization of the protein might permit formation of HB networks between ions and PGs.

## 6 Conclusion

Based on observation of the simulated formation of water HB networks connecting $Ca^{2+}$ and PG oxygens interior to a transient beta sheet involving residues that in a correct fold would not form this sheet, we propose that elimination of these transient misfolded sheets in the course of folding is ion-dependent. Alternatively stated, amino acid preference for beta sheet by sidechain blocking of backbone hydration is not effective unless ions are present. Backbone hydration of these transient beta sheets without ions might occur on longer time scales than simulated, but it appears this hydration is faster in the presence of $Ca^{2+}$ and is anticipated to form closer HB to PG oxygen and to hydrate the backbone where water alone does not.

It is anticipated that alpha helix preferring residues may appear in a beta sheet if surrounding residues compensate for the deficiency of sidechain blocking of backbone hydration, referred to here as sidechain complementation. It is desirable that the sidechain structures of complementation be studied, and we expect it will soon be possible to do this with QMD. Non-hydrophobic residues will also vary the formation of HB networks to PG oxygen, and use of similar methods is indicated for such a study.

The proposed role of cations in eliminating transient secondary structures which arise due to misfolding, an early stage of which is computationally demonstrated here with $Ca^{2+}$, invites consideration of high $Ca^{2+}$ concentrations in the ER as a testing environment for correct sidechain complementation and hence fold. It is suggested that the exposed PG oxygen in an otherwise hydrophobic setting, perhaps in association with a multivalent cation, could mark the hydrophobic patch for action by a folding chaperone.

Limitations due to thin solvent shell, short simulated time span and use of established DFT methods that suffer disturbances in modelling PG resonance and hence RAHB in protein secondary structures [11] are acknowledged. QMD methods with runtime and memory use that scales linearly with the total of protein atom count and solvent atom count, are accurate in their calculation of PG resonance and are applicable through the transition metal block are highly desirable. If such methods are wavefunction, high quality basis sets are necessary for accuracy with proteins [11]. Whereas present simulated time-scales extend to ~1 picosecond, the 1 nanosecond mark might become accessible with such methods. Whereas present solvation depth is 7 angstroms over short sidechains, 20 angstroms over the longest sidechains is desirable so that surface effects of the vacuum/water interface are not experienced at the water/protein interface, and this vast increase in solvent might become possible with such methods.

## 7 Future work

Summarizing the investigations that would be enabled by the availability of the methods with the above characteristics:

- Survey backbone hydration in all RAHB secondary structure types in the presence of each biologically relevant cation, including what combinations of residues are complementary for blocking backbone hydration
- With linear scaling NBO, analyse charge transfer and polarization in HB networks of water connecting PG and ions for each QMD frame
- Survey loss of sidechain complementation in proteins of natural sequence with increasing temperature and ion concentration in mixed ionic environments, validating against data which has been physically obtained
- Observe what happens at longer simulated time scales – is there indication that folding could proceed unassisted by chaperone in the case of direct contact between cation and backbone oxygen? If such contact occurs, attempt binding of molecular chaperones to cation-disrupted secondary structures
- In the more distant development of linear scaling QMD methods that have sufficiently low constant factor on resource use, simulate entire folding of fast folding proteins to observe cation-dependent backbone hydration

The combinatorial nature of these investigations requires many simulations and a total simulation time that makes necessary methods that are linear scaling with the total of protein, solvent and ion atom count.

Ahead of the availability of methods suitable for these surveys, the Protein Data Bank might be surveyed for existence of surface backbone oxygens in a largely hydrophobic surface patch, and what features of the patch might prevent chaperone binding or action considered.

## 8 Acknowledgements

Prof. John A. Carver is acknowledged for reading this manuscript and offering editing suggestions.

eResearch South Australia is acknowledged for hosting and administering machines provided under Australian Government Linkage, Infrastructure, Equipment and Facilities grants for Supercomputing in South Australia and directing funds to the acquisition of Nvidia Tesla GPU nodes.

## 10 Appendix 1. Backbone hydration in experiment 1302

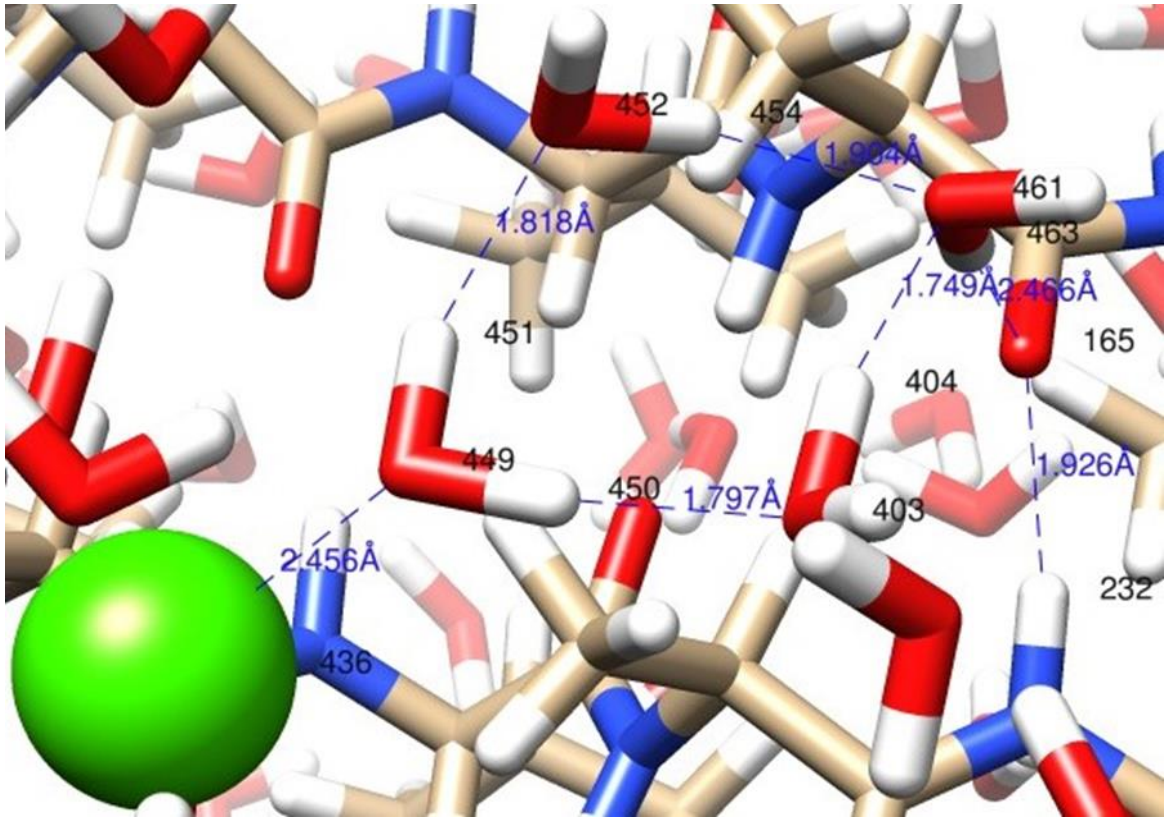

Figure 8. Backbone Hydration Outside Experimental Surface in Final Frame of Experiment 1302

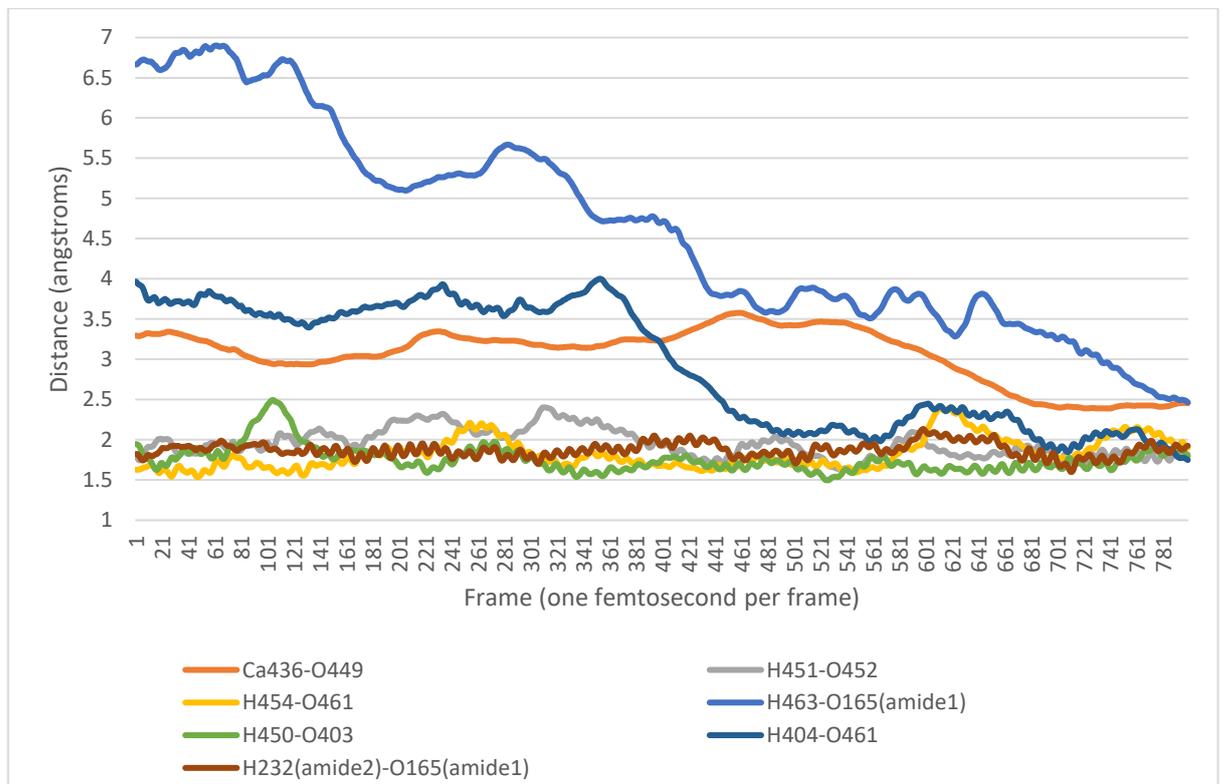

Figure 9. Selected bond lengths in experiment 1302. Atom Ids Shown in Figure 8

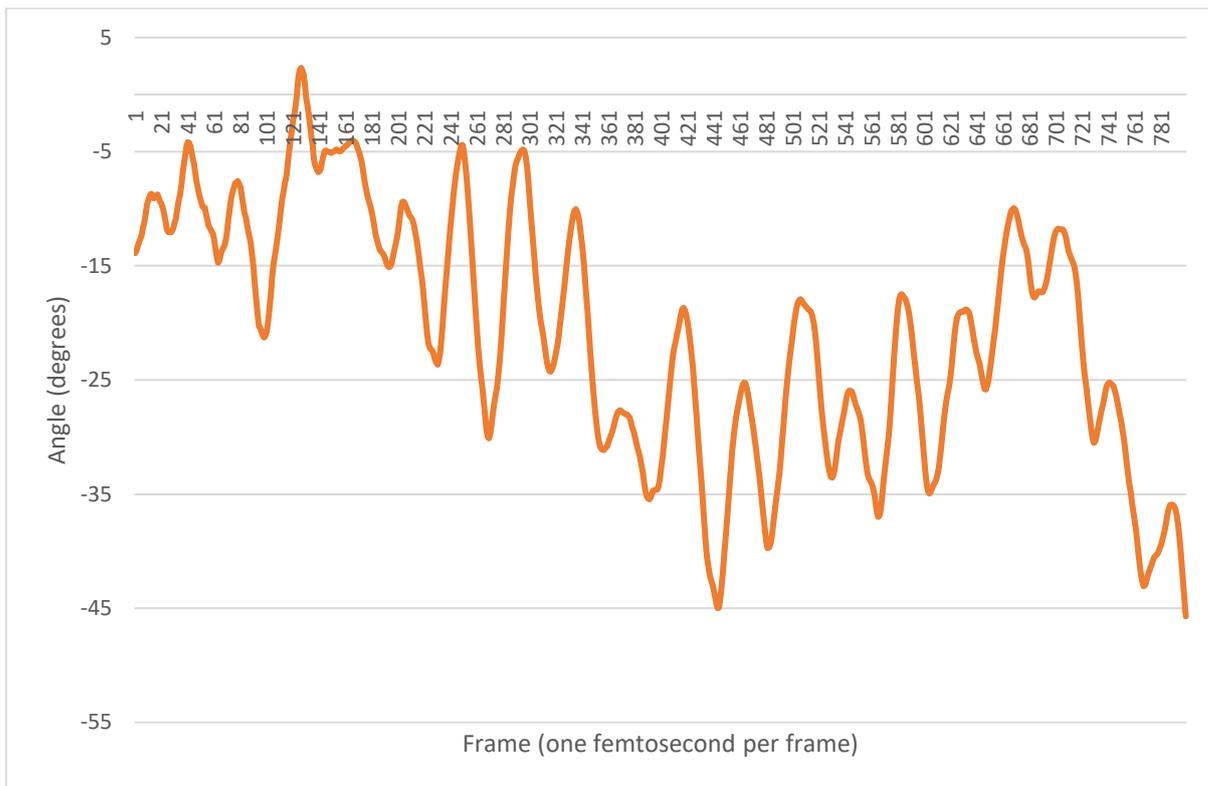

Figure 10. Angle of O165-H232 from O165, C163, N164 Normal Minus 90.0 in Experiment 1302

## 11 Appendix 2. Backbone hydration in experiment 1303

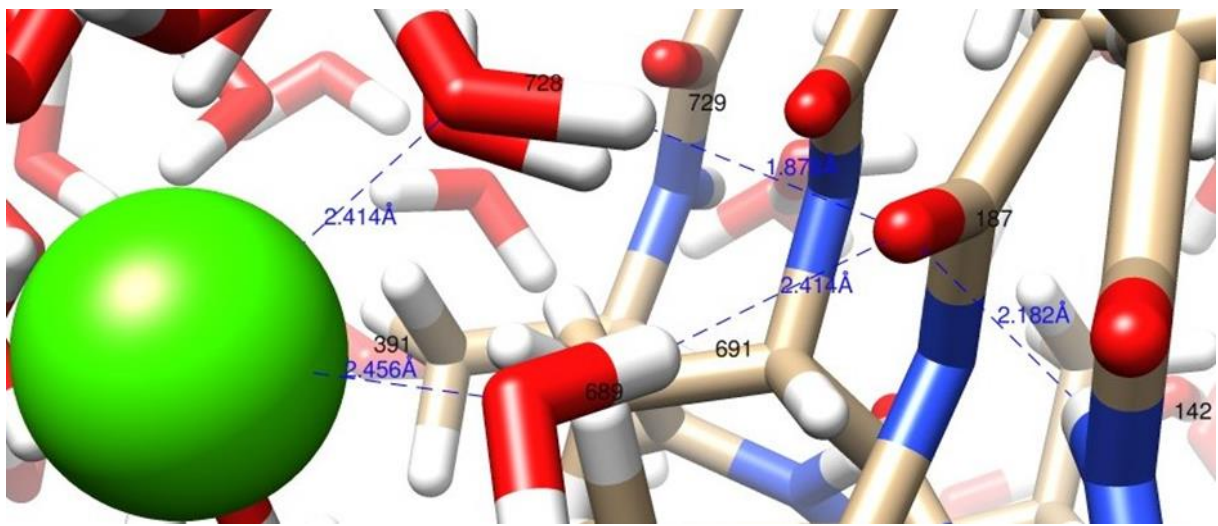

Figure 11. Backbone Hydration Outside Experimental Surface in Final Frame of Experiment 1303

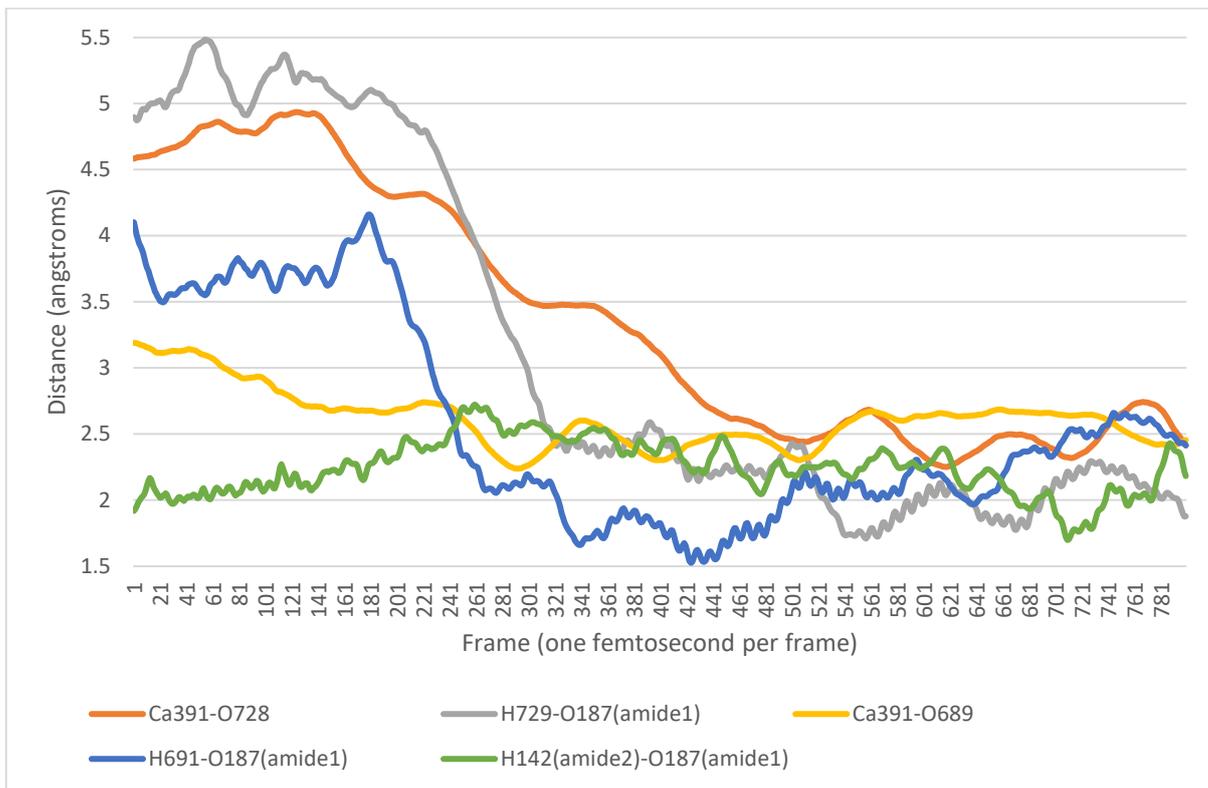

Figure 12. Selected Bond Lengths in Final Frame of Experiment 1303. Atom Ids Shown in Figure 11

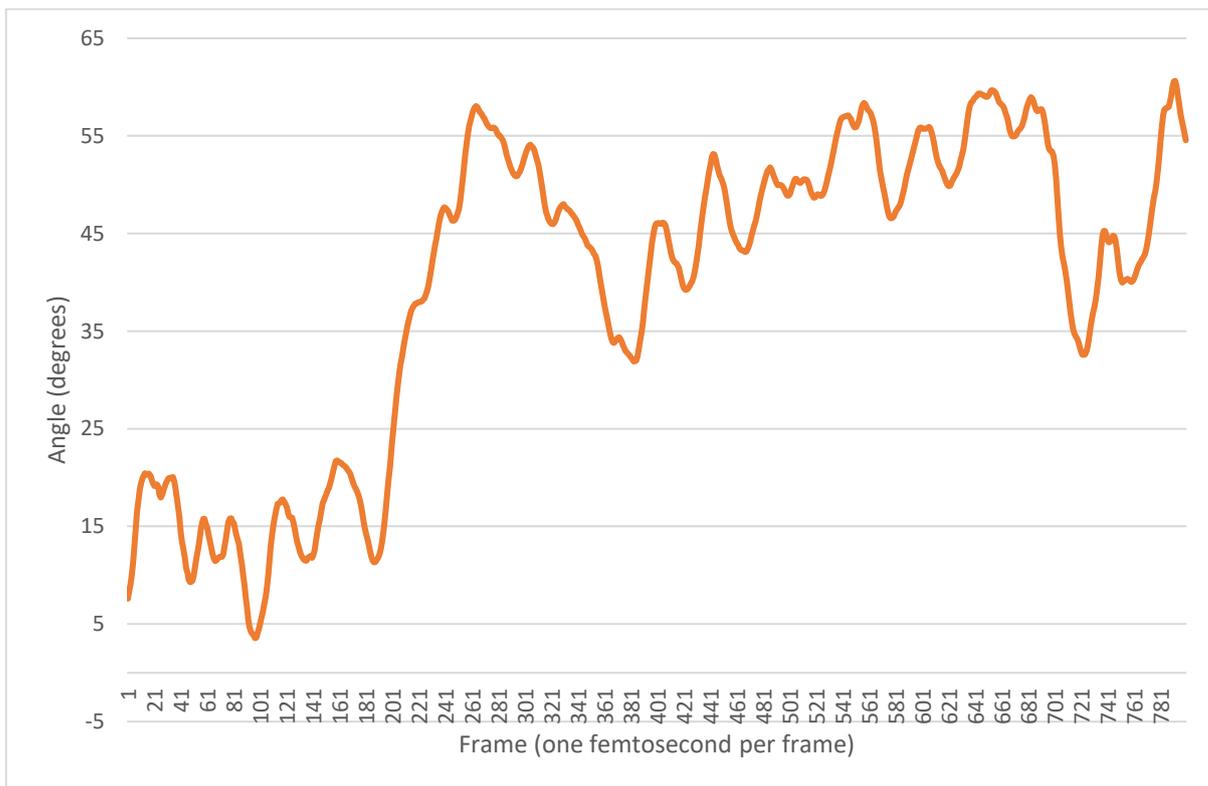

Figure 13. Angle of O187-H142 from O187, C185, N186 Normal Minus 90.0 in Experiment 1303

18